\documentclass[12pt]{iopart}
\usepackage{graphicx}

\def\sech{\mbox{sech}}
\def\sn{\mbox{sn}}
\def\cn{\mbox{cn}}
\def\sd{\mbox{sd}}
\def\nd{\mbox{nd}}
\def\am{\mbox{am}}
\def\ds{\displaystyle}

\begin{document}
\title[Effects of finite curvature on soliton dynamics]
{Effects of finite curvature on soliton dynamics \\
in a chain of nonlinear oscillators}

\author{Peter L Christiansen\dag, 
Yuri B Gaididei\ddag\ and \\ Serge F Mingaleev\ddag}

\address{\dag\ Department of Mathematical Modelling,
The Technical University of Denmark, DK-2800 Lyngby, Denmark}

\address{\ddag\ Bogolyubov Institute for Theoretical Physics,
Kiev 03143, Ukraine}


\begin{abstract}
We consider a curved chain of nonlinear oscillators and show
that the interplay of curvature and nonlinearity leads to 
a number of qualitative effects. In particular, the energy 
of nonlinear localized excitations centered on the bending 
decreases when curvature increases, i.e. bending manifests
itself as a trap for excitations. Moreover, the potential 
of this trap is {\em double-well}, thus 
leading to a {\em symmetry breaking} phenomenon: a symmetric
stationary state may become unstable and transform into 
an energetically favorable asymmetric stationary state.
The essentials
of symmetry breaking are examined analytically for a
simplified model. We also demonstrate a threshold character
of the scattering process, i.e. transmission, trapping, 
or reflection of the moving nonlinear excitation passing 
through the bending. 
\end{abstract}



\section{Introduction}

Recently, determination of the dynamical properties of physical
systems with nontrivial geometry has attracted a growing 
interest
because of their wide applicability in various physical and
biophysical problems. Examples are biological macromolecules,
nanotubes, microtubules, vesicles, electronic and photonic
wave-guide structures \cite{dna,nan,mict,amp,imry,souk}; the
gene transcription is usually accompanied by a local DNA 
bending
\cite{reiss,heu,burl}; the electronic properties of carbon
nanotubes drastically depend on their chirality \cite{hamada};
T-shape junctions were recently proposed \cite{menon} as 
nanoscale
metal-semiconductor-metal contact devices; periodically spaced,
curved optical waveguides were proposed for observation of 
optical Bloch oscillations \cite{lenz}.

The geometry manifests itself in particular in 
creation of linear quasi-bound states (see e.g.
\cite{kircz,gai}). On the other hand, it is well
known that the balance between nonlinearity and dispersion
provides an existence of spatially localized 
soliton-like excitations. Taken together these two localization
mechanisms compete and one may expect that the 
{\em interplay of geometry and nonlinearity can lead 
to qualitatively new effects}.

An example of nonlinearity-induced change of the geometry of
the system can be found in Ref. \cite{dan} where the 
classical Heisenberg model on an infinite elastic cylinder was
considered. It was shown that a periodic
topological spin soliton induces a periodic pinch of the
cylindrical manifold. These results are applicable to
microtubules and vesicles comprised of magnetic organic 
materials \cite{sax}. 
The stationary lasing of the modes in a microdisk was 
investigated in Ref. \cite{taka} where it was shown that 
nonlinear whispering gallery modes exist in the microdisk 
above some finite pumping thresholds.

In this paper we consider a curved chain of nonlinear
oscillators which is described by the nonlinear 
Schr{\"o}dinger (NLS) model with the coupling coefficients 
depending on the distance between oscillators. 
This model has been in particular used as an approximation 
to the nonlinear dynamics of biological macromolecules, 
such as proteins \cite{dav,scott,leonor} and DNA \cite{yakush}.
It was shown that nonlinear excitations should play a key 
part in the biological functioning of such molecular chains.
However, usually they are investigated on the assumption 
that the chain has straight rod-like configuration. 
On the other hand, thermodynamical properties of these 
macromolecules cannot be understood without taking into
account their flexibility and existence of non-trivial
spatial configurations. Moreover, there is increasing
evidence that non-trivial spatial configurations should 
play a part in nonlinear dynamics of the macromolecules. 
In particular, the DNA molecule is usually hooked 
\cite{reiss,heu,burl} by the RNA-polymerase which binds 
to the promoter segment of DNA and then initiates the 
DNA transcription. 

This raises the interesting question of whether a 
non-trivial spatial configuration of the chain could 
introduce some new qualitative effects into the nonlinear 
dynamics of the system? In the present paper we demonstrate 
some of such effects on the example of the chain with 
a single parabolic bending. In Sec. 2 we describe the 
model and present the results of numerical calculations of
stationary excitations. In particular, we discuss the 
multistability phenomenon and curvature-induced symmetry 
breaking in the spectrum of nonlinear excitations. 
In Sec. 3 we analyze analytically a minimal model which 
gives qualitatively the same behavior as the original 
model. In Sec. 4 we consider dynamical properties of 
nonlinear excitations, in particular scattering of the
moving excitations by the chain bending. 
Section 5 presents the concluding discussion.

\section{System and equations of motion}

\begin{figure}
\centerline{
\includegraphics[width=65mm,clip]{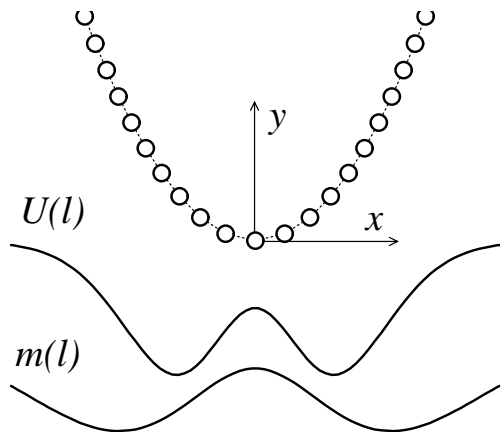}}
\caption{The curved chain under consideration and the
corresponding effective double-well potential $U$ and 
mass $m$, created by the bending.}
\label{fig:chain}
\end{figure}

The model we study is described by the Hamiltonian
\begin{eqnarray}
\label{ham}
H = \sum_{n} \left\{ 2 |\psi_n|^2 - \sum_{m \neq n} 
J_{n, m} \psi^*_n \, \psi_m - \frac{1}{2} |\psi_n|^4 
\right\} \; ,
\end{eqnarray}
where $\psi_n(t)$ is the complex amplitude at the site $n$
($n=0,\pm 1,\pm 2$, \ldots).  For instance in the case of the 
two-strand DNA model \cite{triest}, $\psi$ is the complex 
amplitude of the base-pair stretching vibration. 
We assume that the excitation transfer $J_{n, m}$
in the dispersive term depends on the distance in the embedding
space between the sites $n$ and $m$:  $J_{n, m} \equiv
J(|\vec{r}_n-\vec{r}_{m}|)$, where the radius-vector
$\vec{r}_n=(x_n, y_n, z_n)$ characterizes spatial 
position of the site $n$. Of course, when $J_{n,m}$ models 
electromagnetic coupling between dipoles, this coefficient will 
depend also on the relative angles. However, in some cases
such dependence is not present. E.g., for the semiflexible 
ribbon polymers \cite{golestanian} the dipoles are
orthogonal to the plane in which the chain can be curved,
so that the corresponding interaction would depend only 
on the distance between the dipoles.

From the Hamiltonian (\ref{ham}) we obtain the equation of
motion in the form
\begin{eqnarray}
\label{nls}
i \frac{d}{dt} \psi_n - 2 \psi_n +\sum_{m \neq n} 
J_{n, m} \psi_m + |\psi_n|^2  \psi_n=0 \; .
\end{eqnarray}
The Hamiltonian $H$ and the number of excitations (quanta)
$N=\sum_n|\psi_n|^2$ are both conserved quantities. The sites
are assumed to be equidistantly placed on a planar envelope 
curve.

We consider the properties of nonlinear excitations in the
vicinity of the bends where the envelope curve can be 
modeled by a parabola (see the top part of
Fig. \ref{fig:chain}):  $y_n=\kappa\,x^2_n/2$ and $z_n=0$ with
$\kappa$ being the inverse radius of curvature at the bending
point. We assume that the chain is inextensible that is 
the distance between neighboring sites is constant: 
$|\vec{r}_{n+1}-\vec{r}_n|=1$ for all $n$ and $\kappa$. 
Therefore, when $\kappa$ is not
too large ($|\vec{r}_{m}-\vec{r}_n| > 1$ for all $|n-m| > 1$)
the excitation dynamics does not depend on the curvature of the
system in the {\em nearest-neighbor approximation}, when
$J_{n, m}=J\delta_{n-m, \pm 1}$. 

\begin{figure}
\centerline{
\includegraphics[width=100mm,clip]{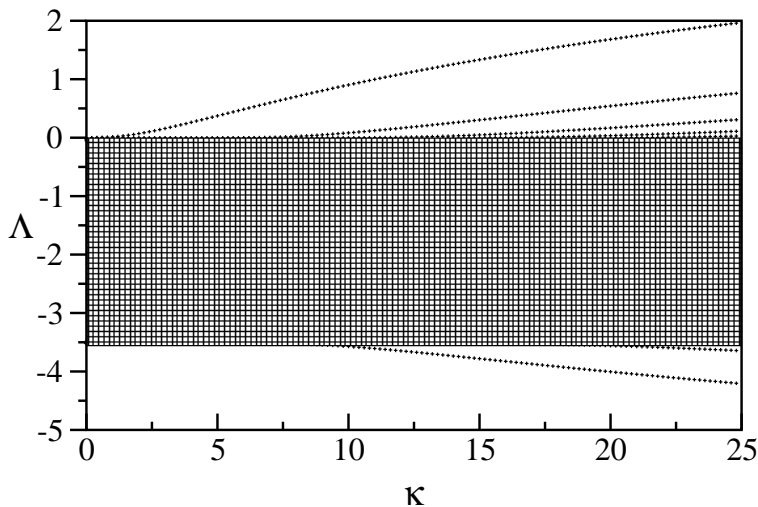}}
\caption{Dependence of the spectrum of linear excitations 
on the curvature of the chain for $\alpha=2$ and $J=6.4$.
Localized linear bound states emerge above 
(symmetric states) and below (antisymmetric states) 
the continuum spectrum of extended states.}
\label{fig:lin-spect}
\end{figure}
\begin{figure}
\centerline{
\includegraphics[width=100mm,clip]{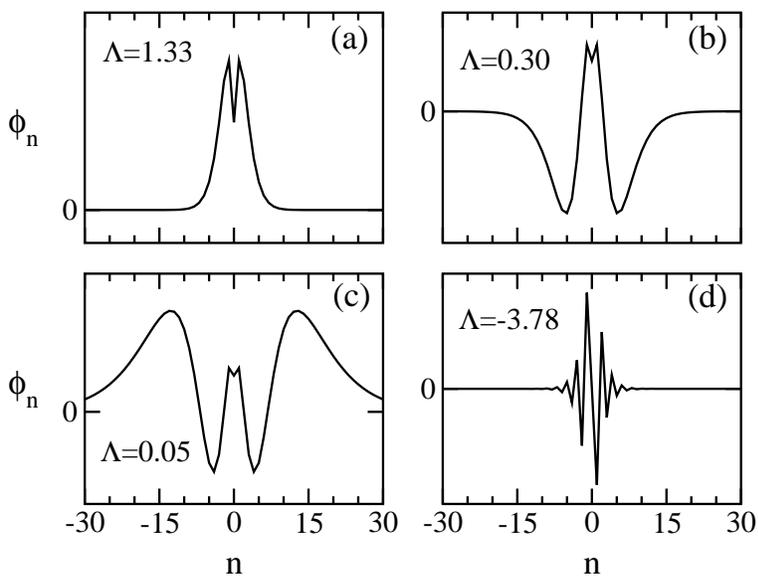}}
\caption{Profiles of the linear bound states from Fig. 
\protect\ref{fig:lin-spect} for $\kappa=15$.}
\label{fig:lin-modes}
\end{figure}

In many physical problems,
however, the nearest-neighbor approximation is too crude. 
For example, effective dispersive interaction between nonlinear layers 
in nonlinear dielectric superlattices is exponential, 
$\exp(-\alpha r)$, with the inverse radius $\alpha$ depending on 
parameters of the lattice (superlattice spacing, linear refractive 
index and so on) \cite{sup}.
In the framework of the Peyrard-Bishop model \cite{pb,dpb} 
the nonlinear dynamics of the DNA molecule can be described by 
Eq. (\ref{nls}) with nonlinear term of the opposite 
sign (because of the {\em repulsive} anharmonicity). In this case 
$\psi$ is the complex amplitude of the base-pair stretching 
vibration \cite{triest}. The base-pairs (AT, GC) in the DNA molecule 
are usually asymmetric (see, e.g., \cite{dna}), and therefore
the vibration which corresponds to the stretching of the hydrogen 
bonds connecting the two bases, transfers along the chain due to 
the dipole-dipole interaction decaying with the distance $r$ as 
$1/r^3$. Since the dipole moments of base-pairs are almost 
perpendicular to the molecular axis \cite{dna}, they are almost 
parallel to each other and the interaction between dipoles is 
also repulsive ($J_{n, m}<0$). Thus, after substitutions $t \to -t$ and 
$J_{n, m} \to -J_{n, m}$, this model of the DNA molecule will be 
described by Eq. (\ref{nls}). 
We have investigated both cases,
$J_{n, m}= J \exp\left(-\alpha |\vec{r}_n - \vec{r}_{m}| \right)$
and $J_{n, m}= J |\vec{r}_n-\vec{r}_{m}|^{-s}$,
with qualitatively the same results.
In this paper, however, we discuss only the first case in
detail. It is known \cite{magn} that in the case of straight
rod-like chain ($\kappa=0$) such NLS model exhibits 
bistability in the spectrum of nonlinear stationary states 
for $\alpha < 1.7$. 
To distinguish these effects from the finite curvature
effects we use $\alpha=2$ in what follows.

For stationary states, 
$\psi_n(t)=\phi_n(\Lambda) \, e^{i\Lambda t}$, where
$\Lambda$ is the nonlinear frequency and $\phi_n(\Lambda)$ 
is the excitation amplitude of the $n$-th site, 
Eq. (\ref{nls}) reduces to the system of nonlinear algebraic
equations 
\begin{eqnarray}
\sum_{m \neq n} J_{nm} \phi_m 
+ \phi_n^3 = (2+\Lambda) \, \phi_n \; ,
\label{eq-nls-stat}
\end{eqnarray}
which can be solved, for instance, by the Newton-Raphson
iterations. 

In Fig. \ref{fig:lin-spect} we demonstrate that 
in the linear limit, i.e. when the term 
$\phi_n^3$ in Eq. (\ref{eq-nls-stat}) is neglected, the
bending creates at least one linear localized state. 
The profile of the first bound state, which is always present, 
is shown in Fig. \ref{fig:lin-modes}a. When the 
curvature of the chain increases, new localized states 
are emerged from the 
continuum spectrum; their profiles are shown in 
Fig. \ref{fig:lin-modes}. As it has been recently proven 
for another but similar system \cite{Exner},
the number of these bound states can exceed any positive 
integer provided the chain is curved ``enough''.

\begin{figure}
\centerline{
\includegraphics[width=100mm,clip]{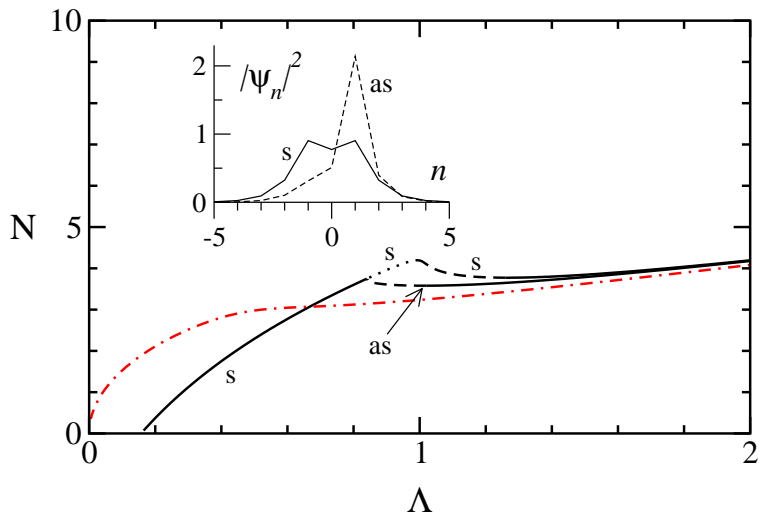}}
\caption{Number of excitations $N(\Lambda)$ for the straight
chain ($\kappa=0$ ; {\em dot-dashed line}) and curved chain
($\kappa=3$ ; {\em solid lines} for stable states, 
{\em dashed lines} for the states unstable with respect to
the breathing mode, and {\em dotted line} 
for the states unstable with respect to the Peierls mode)
for $\alpha=2$ and $J=6.4$.
In the inset shapes of the symmetric (s) and asymmetric (as)
stationary states for $N=3.6$ are shown.}
\label{fig:norm}
\end{figure}

However, in the present paper we shall study the case
of smooth bendings for which only the first bound state 
is present. In Fig. \ref{fig:norm} we plot the dependence 
of $N(\Lambda)$ obtained numerically for both, the straight 
rod-like and curved chains. As may be inferred from 
this figure, several new features arise as a consequence 
of the finite curvature.

\begin{figure}
\centerline{
\includegraphics[width=80mm,clip]{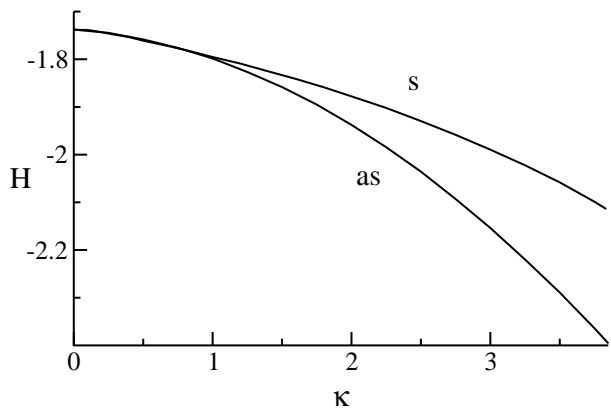}}
\caption{Energy of symmetric (s) and asymmetric (as) nonlinear
excitations versus the curvature $\kappa$ for $\alpha=2$,
$J=6.4$, and $N=4$.}
\label{fig:energy}
\end{figure}

Specifically, {\em there appears a gap} 
at small $\Lambda$ in which no 
localized solution can exist. As can be seen from Fig. 
\ref{fig:lin-spect}, this gap originates from 
the existence of the linear two-hump bound state discussed 
above. 

In the case of curved chain there are {\em two 
branches of stationary states}: a branch of symmetric 
localized excitations (s), which exists for all values
of $N$, and a branch of {\em asymmetric localized excitations}
(as), which exists only for $N > N_{th}(\kappa)$.  The threshold
value $N_{th}(\kappa)$ decreases when $\kappa$ decreases and
vanishes in the limit $\kappa \to 0$.
The symmetric stationary state has a two-humped shape
(evolved from the linear two-hump localized mode)
while in the asymmetric state the maximum is shifted either
to the left or to the right from the center of symmetry $n=0$
(see inset in Fig. \ref{fig:norm}).

In contrast to the case of straight rod-like chain 
(see dot-dashed line in Fig. \ref{fig:norm}), the dependence
$N(\Lambda)$ for the symmetric stationary states in curved 
chain is {\em non-monotonic}. That is, there exist an 
interval of $N$ in which three different symmetric states 
coexist for each excitation number; two of them are 
usually stable \cite{magn,lr}. Such bistability phenomenon 
exists in the straight rod-like chain only for smaller 
values of $\alpha<1.7$. Thus, one may conclude that the 
bending of the chain facilitates the {\em bistability of 
nonlinear excitations}.

The energy of the symmetric as well as asymmetric localized
states is monotonically decreasing function of the curvature
(see Fig. \ref{fig:energy}). Thus, one may expect that the 
{\em nonlinear localized excitation may facilitate bending 
of flexible
molecular chain}.  For $N > N_{th}(\kappa)$, when symmetric and
asymmetric stationary states coexist, the {\em asymmetric state
is always energetically more favorable}.  Taking into account
that in the linear limit the ground state of the model is
symmetric with respect to the center of symmetry $n=0$ one can
conclude that {\em combined action of the finite curvature and
nonlinearity provides the symmetry breaking} in the system.

Apart from the smooth bending we studied the case of the wedge:
$y_n=|x_n| \,\tan(\theta) \, $, and obtained qualitatively the
same results.  We checked also that the exact form of the
dispersive interaction is not of serious concern here. To this
end we studied the case when the dispersive interaction has the
power dependence on the distance:
$J_{n, m}= J |\vec{r}_n-\vec{r}_{m}|^{-s}$.
Qualitatively the same results were obtained for this case, too.
The symmetry breaking, which we observe, means that there is a 
bistability in the spectrum of nonlinear excitations: the excitation 
can be localized either to the left or to the right of the 
bending. By applying an external periodic field one can 
achieve switching between these two states. In this way we 
may expect a new resonance to appear in the response
of the system to a periodic driving.

\section{Analytical approach and minimal model}

To gain a better insight into the symmetry breaking 
mechanism, let us consider the continuum limit of the 
discrete NLS model given by Eqs.  (\ref{ham})--(\ref{nls}). 
We are interested here in the case when the characteristic 
size of the excitation is much larger than the lattice 
spacing.  It permits 
us to  replace $\psi_n(t)$  by the function $\psi(\ell,t)$ of the 
arclength $\ell$ which is the continuum analogue of $n$. Using the
Euler-Mclaurin summation formula \cite{abr} we get
\begin{eqnarray}
\label{hamc}
H=\int\limits_{-\infty}^{\infty} \Biggl\{ \frac{1}{2 \, m(\ell)}
\Bigl| \frac{\partial \psi}{\partial \ell} \Bigr|^2+
U(\ell)|\psi|^2 - \frac{1}{2} |\psi|^4 \Biggr\} d \ell \; .
\end{eqnarray}
Here
$m^{-1}(\ell)= \int\limits_{-\infty}^{\infty} 
(\ell'-\ell)^2\,J_{\ell, \ell'} d \ell'$ 
is the inverse effective mass of the excitation, and 
$U(\ell) = -\int\limits_{-\infty}^{\infty}\, 
J_{\ell, \ell'} d \ell'$ 
is the energy shift due to the dispersive interaction. 
Near the bending point the number of neighbors available for 
excitation is larger than far away where the curvature is small. 
Near the bending the effective potential, $U(\ell)$,
exhibits a double-well shape. This fact together with the resulting
double-well spatial dependence of the effective mass, $m(\ell)$, are 
the reasons why the excitation
is localized. The attractive potential $U(\ell)$ 
consists of two symmetric wells whose positions ($\,\pm\,a$) and 
depths ($\Delta \equiv U(|\ell|\rightarrow\infty)-U(a)$) are 
determined by the range of the dispersive interaction and the 
peculiarities of the bending geometry (see Fig. \ref{fig:chain}). 
E.g., in the case of the wedge, $y_n=|x_n|\,\tan (\theta) \,$, 
we get for $\theta\ \,< \,\pi/4$
\begin{displaymath}
U(\ell)=-\frac{J}{\alpha}\,(2-e^{-\alpha|\ell|}) - 
\frac{J}{\alpha} \left\{
\begin{array}{ll}
\,\ds{e^{-\alpha|\ell|\cos(2\theta)}}
&\textrm{when $~\theta\,<\,1/2 \; ,$}\\
\alpha\,|\ell|\,K_1(\alpha|\ell|) & \textrm{when
$\theta\,\rightarrow \pi/4 \; ,$}
\end{array}\right.
\end{displaymath}
where $K_n(z)$ denotes the modified Bessel function \cite{abr}.
The distance between the potential minima 
$2 a \approx (1+\theta^2)/\alpha$ 
and their depth 
$\Delta \approx 4\,e\,J\,\theta^2/\alpha$
increase when the wedge angle $\theta \rightarrow \pi/4$.
Note that the Hamiltonian similar to Eq. (\ref{hamc}) with 
$U(\ell) \equiv 0$ was introduced in Ref. \cite{peyrard} to 
study the effects of the DNA bending on breather trapping. The 
step function dependence of the effective mass $m(\ell)$ was 
there assumed.

\begin{figure}
\centerline{
\includegraphics[width=100mm,clip]{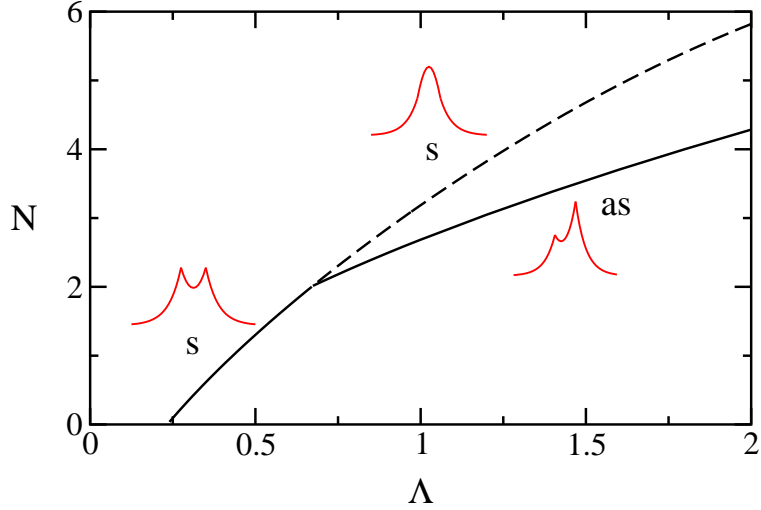}}
\caption{The dependence $N(\Lambda)$ for symmetric (s) and asymmetric
(as) nonlinear excitations in the two-impurity minimal model
(\protect\ref{toy}) with $\epsilon=0.7$ and $a=1$.}
\label{fig:toy}
\end{figure}

The NLS model (\ref{hamc}) is still very difficult to solve 
analytically.   But it is interesting anyway to find a model --- 
chosen as simple as possible --- which gives qualitatively the 
same results as the original model. Such a model can be obtained by 
using step-wise approximation
\begin{eqnarray}
\label{stp}
U(\ell)=-\frac{2J}{\alpha} - \Delta 
\sum_{\sigma=\pm} \left( H(\ell-\sigma a+w/2) - 
H(\ell-\sigma a-w/2) \right) \; , 
\end{eqnarray}
where $H(x)$ is the Heaviside step-function, $w$ is the width of 
the potential well. The similar expression can be written for the 
effective mass $m(\ell)$. In the step-wise approach the 
stationary equation which corresponds to the Hamiltonian 
(\ref{hamc}) can be solved and the corresponding eigenvalue 
problem can be explored. However,results obtained in this manner 
look very cumbersome. Therefore we shall make a further 
simplification by applying the limit $w \rightarrow 0$, keeping 
the product $~w\,\Delta\equiv\epsilon$ constant in Eq. 
(\ref{stp}). In this way we arrive at the following minimal model
\begin{eqnarray}
\label{toy}
i\,\frac{\partial \psi}{\partial t} +
\frac{\partial^2 \psi}{\partial \ell^2} +
\epsilon (\delta(\ell-a)+\delta(\ell+a)) 
\psi + |\psi|^2 \psi=0 \; ,
\end{eqnarray}
where the gauge-transformation 
$\psi\rightarrow\psi \exp\left(-i2Jt/\alpha\right)$ was used.
Introducing into Eq. (\ref{toy}) the stationary state ansatz
$\psi(\ell,t)=\phi(\ell) \exp(i\Lambda )$ 
where $~\phi(\ell) \rightarrow 0~$ when 
$~|\ell| \rightarrow \infty~$, we find that
\begin{displaymath}
\fl 
\phi(\ell)= \left\{
\begin{array}{ll} 
\sqrt{2\Lambda} \, \sech \left( \sqrt{\Lambda} \, 
(\ell-\ell_{-1}) \right) & \textrm{when $~\ell\,<\,-a,$}\\
\ds{\sqrt{\frac{2 \Lambda\,(1-m)}{2-m}} \, \nd 
\left(\sqrt{\frac{\Lambda}{2-m}}\, (\ell-\ell_0) \mid m\right)} 
& \textrm{when $|\ell|\,<\,a,$}\\ 
\sqrt{2\Lambda} \, \sech\left(\sqrt{\Lambda}\, 
(\ell-\ell_{1}) \right) 
& \textrm{when $~\ell\,>\,a$}
\end{array} \right.
\end{displaymath}
where $\mbox{pq}(u\mid m)~~(\mbox{p, q = c, s, d, n})$ are the 
Jacobi elliptic functions  with the modulus $m$ \cite{abr}.  The 
parameters $\ell_{\pm 1},\ell_0,$ and $m$ are determined from the 
relations 
\begin{eqnarray}
\label{s1}
\fl
\, \sech \left( \sqrt{\Lambda}\,( a \mp \ell_{\pm 1}) \right) \,
= \sqrt{\frac{1-m}{2-m}} \, \nd \left( \sqrt{\frac{\Lambda}{2-m}} 
(a\mp \ell_0 \mid m) \right), \nonumber\\ 
\fl 
F(\ell_0,m) \equiv \sqrt{1-m\,(1-m)\, \sd^2(u_{+}\mid m )} + \, 
m \, \sd(u_{+} \mid m )\, \cn(u_{+}\mid m ) - \nonumber\\
\fl 
\sqrt{1-m\,(1-m) \, \sd^2(u_{-}\mid m )} + \, m \, 
\sd(u_{-}\mid m) \, \cn(u_{-}\mid m )=0, \nonumber\\ 
\fl 
\sqrt{1-m\,(1-m)\, \sd^2(u_{\pm}\mid m )} + \, m \, 
\sd(u_{\pm}\mid m ) \, \cn(u_{\pm}\mid m)= 
\sqrt{\frac{2-m}{\Lambda}}\,\epsilon \; , 
\end{eqnarray}
where
$u_{\pm}\,=\,\sqrt{\Lambda/(2-m)} \, (a\,\pm \,\ell_0)$. 
Using the above expressions one can evaluate the number of 
excitations which corresponds to the stationary solution
\begin{eqnarray}
\label{numb}
N\,=4\,(\sqrt{\Lambda}-\epsilon)+
\,2\,\sqrt{\frac{\Lambda}{2-m}} \sum_{\sigma=\pm 1}
\,E(\am(u_{\sigma})\mid m) 
\end{eqnarray}
where $\am(u)$ is the amplitude and $E(\phi\mid m)$ is the 
elliptic integral of the second kind \cite{abr}.

The equation $F(\ell_0,m)=0$ always has a  solution $\ell_0=0$. 
This corresponds to the symmetric two-hump wave function 
$\phi(\ell)$. In addition,  $\ell_0\neq 0$ solutions appear for 
$\Lambda \geq \Lambda_{c}$ with the threshold value $\Lambda_c$  
being determined from the condition 
$\partial F / \partial \ell_0=0$ for $~\ell_0\,=\,0$. In other 
words, a symmetry-breaking bifurcation to a doubly degenerate 
branch of asymmetric solutions occurs in the point 
$\Lambda=\Lambda_c$.  It follows from Eq. (\ref{s1}) that the 
symmetry-breaking bifurcation is determined by the equation 
$(1-2 \sn^2 + m \,\sn^4)\,\sqrt{1-2 m \,\sn^2 +m^2 \,\sn^2}
=(1-m) \,\sn \, \cn$ with 
$\sn\equiv \sn\left(\sqrt{\Lambda/(2-m)}\mid m\right)$.

The results of the analytical consideration of the nonlinear 
eigenvalue  problem based on Eqs. (\ref{s1}) and (\ref{numb}) are 
shown in Fig. \ref{fig:toy}. As in the original model the 
stationary state (s) is unique and symmetric for small number of 
quanta $N$. But when the number of quanta exceeds some critical 
value there are two stationary states: symmetric (s) and 
asymmetric (as), with the asymmetric state being energetically 
favorable. Thus, similarly to the original model, the minimal 
model demonstrates a symmetry breaking effect (localization in 
one of the wells). It is worth noting that 
a closely related phenomenon was observed in Ref. 
\cite{akh} where nonlinear electromagnetic waves in
a symmetric planar waveguide were studied. But in 
contrast to our case the nonlinear 
Schr{\"o}dinger model with a step-wise nonlinearity was 
considered.

\section{Dynamical properties of excitations}

\begin{figure}
\centerline{
\includegraphics[width=120mm,clip]{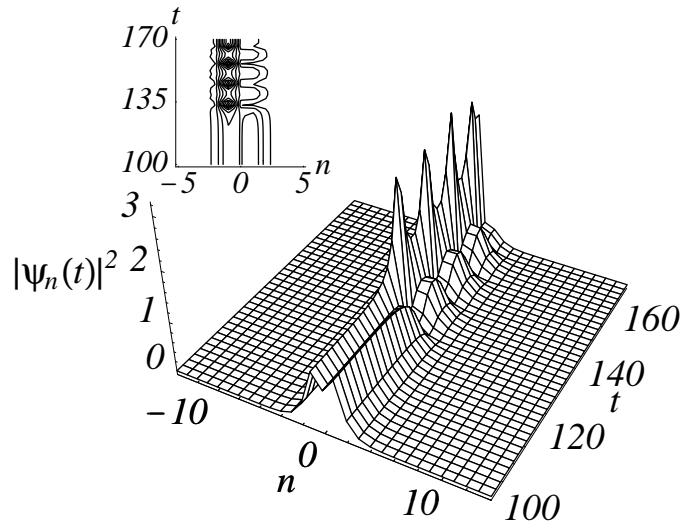}}
\caption{Switching from the symmetric state to the 
asymmetric state at
$\alpha=2$, $J=6.4$, $\kappa=4$, and $N=3.96$.}
\label{fig:dynamics}
\end{figure}
\begin{figure}
\centerline{
\includegraphics[width=0.65\textwidth,clip]{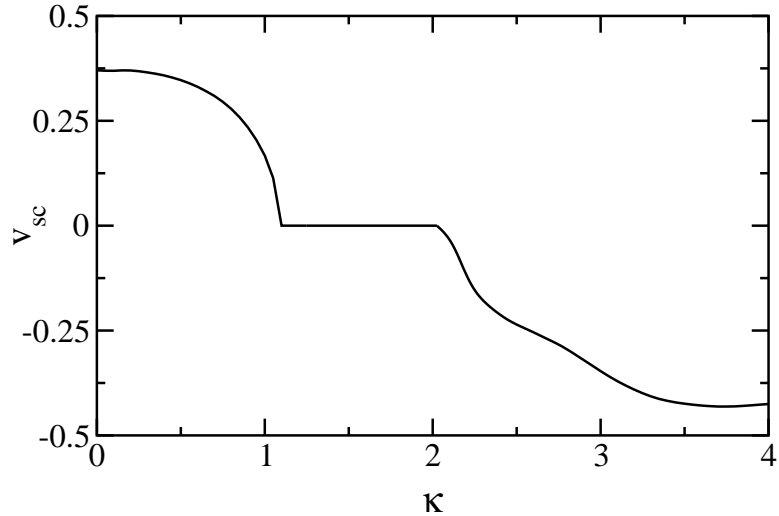}}
\caption{Velocity $v_{sc}$ of the moving part of the excitation 
after the passage of the center of the parabolic chain versus 
the curvature $\kappa$. Here $\alpha=2$, $J=6.4$, $N=2.39$ and 
initial velocity $v=0.37$.}
\label{fig:scatt-v}
\end{figure}


The dynamical properties of nonlinear excitations include 
both, the behavior of immobile stationary excitations 
when they lose stability, and a scattering of the moving 
excitations by the bending.

The first issue has been considered in details in Ref. 
\cite{prl} where we have shown that the symmetric 
two-humped stationary state can lose stability in two ways, 
due to softening of the antisymmetric Peierls internal 
mode and due to softening of the symmetric breathing 
mode.
A typical evolution of the symmetric stationary state in 
the case when it becomes unstable with respect to the Peierls 
mode is shown in Fig. \ref{fig:dynamics}. One can see that the 
two-humped initial symmetric state evolves into an asymmetric 
state with an excited breathing internal mode.


In this paper we consider in more details another item 
which is very intriguing from the viewpoint of possible 
applications: the scattering of moving excitations in 
their passage through the center of the bending. 
To accomplish it, we first obtained a stationary moving 
excitation by pushing (i.e., 
taking $\psi_n(0)=\phi_{n}(\Lambda) \, e^{i k n}$ 
with some $k$ and $\Lambda$) of the initially 
unmoving stationary state $\phi_n$ found numerically 
for the straight chain as described in Sec. 2.
Eventually, evolving over the course of 5000 time 
units, the excitation reached almost stationary form.
Substituting it into
the curved chain far away from the bending, we found out 
that the destiny of the excitation in passing though the 
bending region is strongly dependent on the 
curvature $\kappa$ of the chain. 

\begin{figure}
\centerline{
\includegraphics[width=70mm,clip]{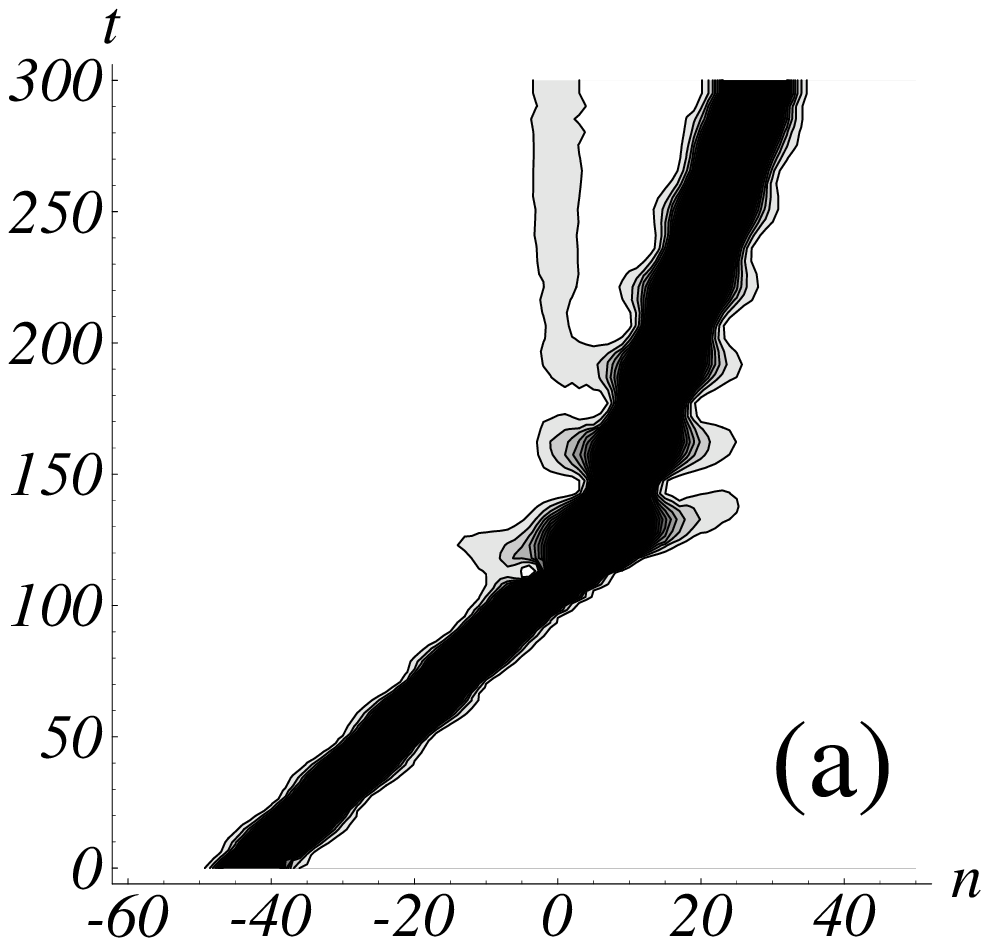}
\hspace{2mm}
\includegraphics[width=70mm,clip]{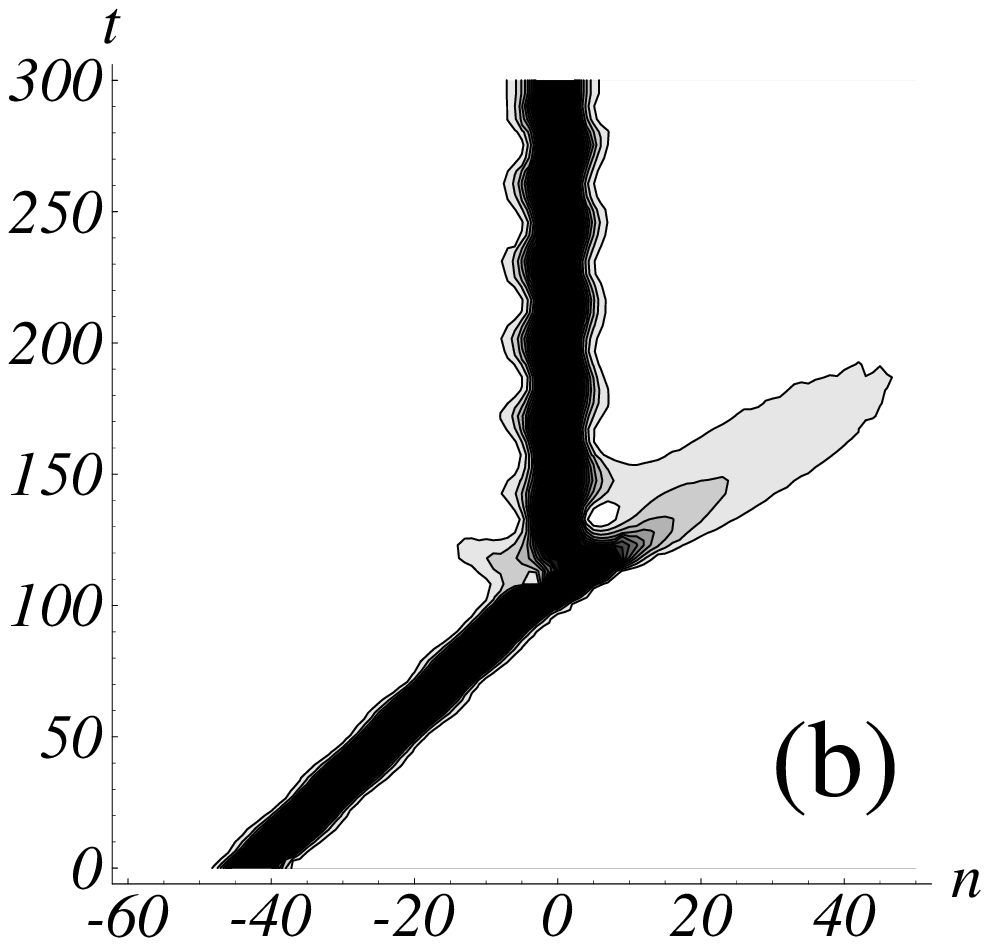}}
\vspace{2mm}
\centerline{
\includegraphics[width=70mm,clip]{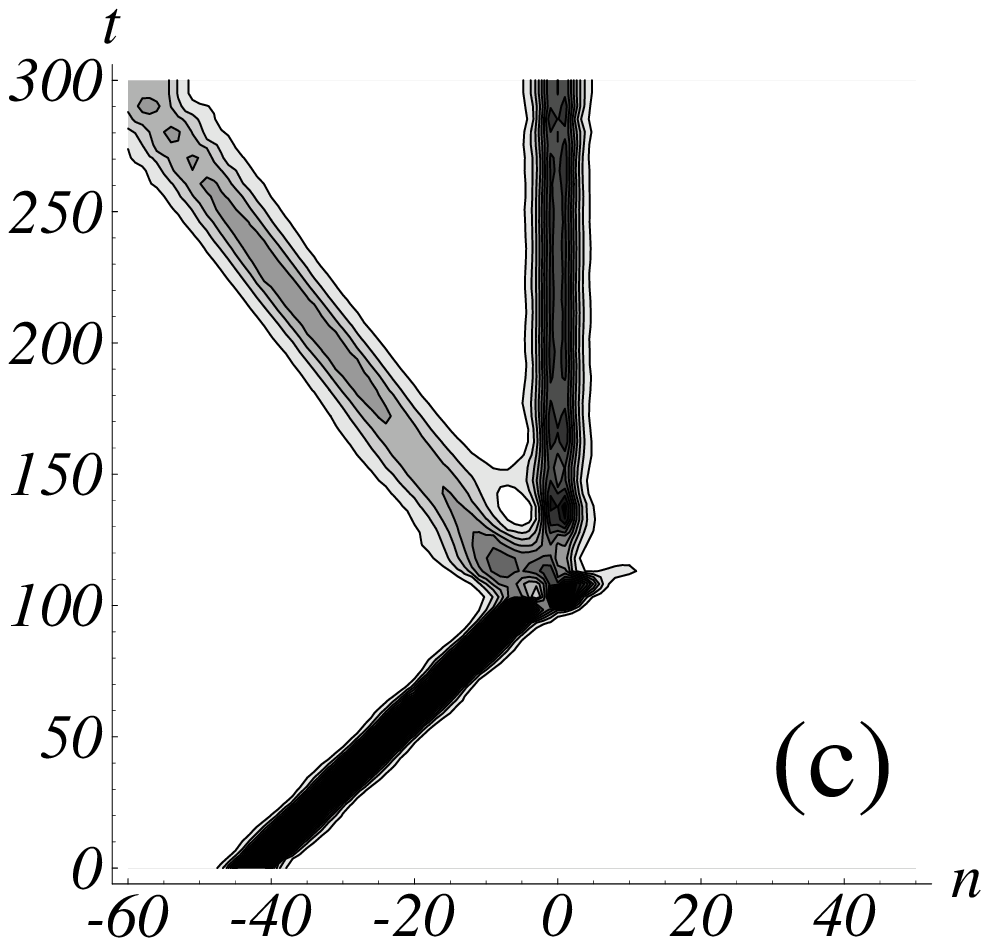}
\hspace{2mm}
\includegraphics[width=70mm,clip]{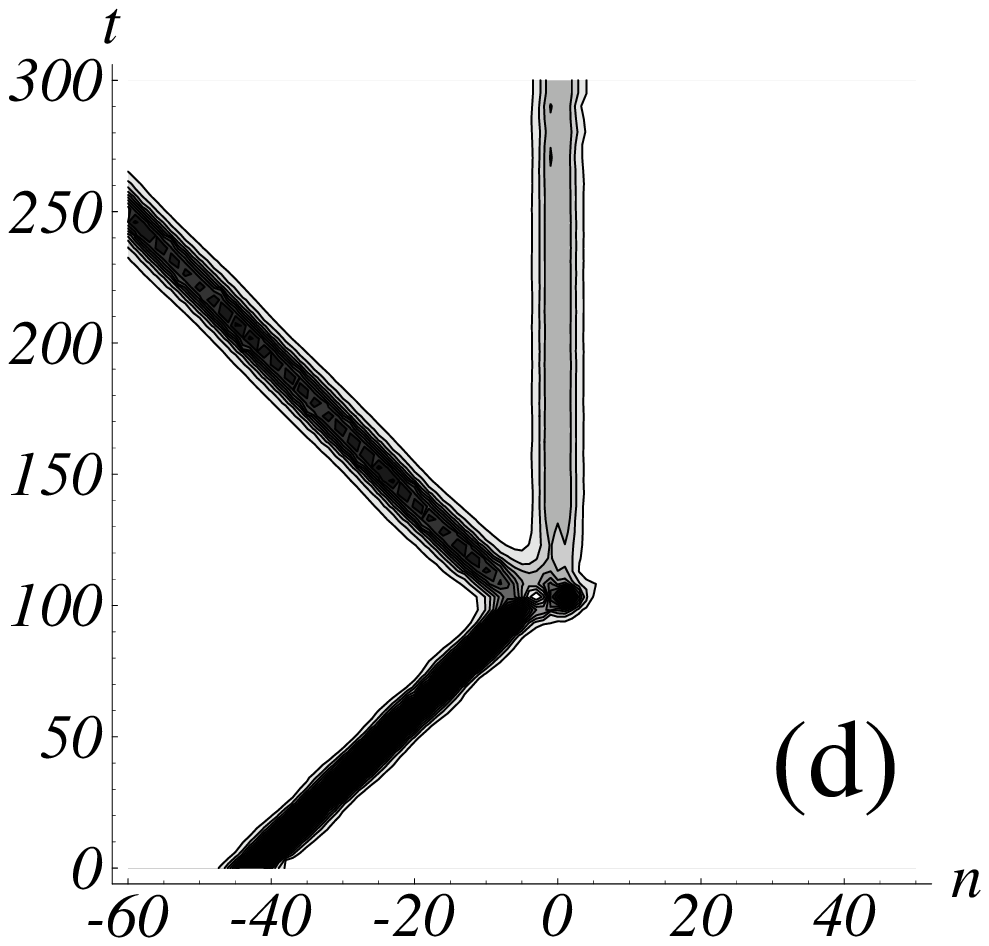}}
\caption{Density plot of $|\psi_n(t)|^2$ which demonstrates 
scattering of the moving excitation from 
Fig. \protect\ref{fig:scatt-v} for different values 
of $\kappa$: (a) 1.05, (b) 1.5, (c) 2.5, and (d) 3.0.}
\label{fig:scatt-1}
\end{figure}

In Fig. \ref{fig:scatt-v} 
we plot the eventual velocity $v_{sc}(\kappa)$ of the 
excitation with $N=2.39$ and initial velocity $v=0.37$. 
One can see that for small $\kappa$ ($\kappa<0.2$) the 
bending does not practically change the excitation 
velocity. However, for $0.2<\kappa<1.07$ the transmission 
becomes clearly inelastic, leading to decreasing of the 
excitation velocity. The excitation passes the 
bending region leaving behind a part of its energy for exciting
of the linear bound state (see Fig. \ref{fig:scatt-1}a).

In the interval $1.07<\kappa<2.1$ the excitation is 
captured by the bending (see Fig. \ref{fig:scatt-1}b), so
that in this case the bending clearly manifests itself as 
an effective trap. Similar phenomenon has been previously 
considered \cite{peyrard} as an effective breather 
trapping mechanism for DNA transcription in the case of
curved DNA. However, in Ref. \cite{peyrard} the bending has
been accounted effectively, postulating some modification 
of the coupling constants around it.

Finally, for $2.1 < \kappa <2.6 $ the excitation is 
reflected from the bending, however a significant part 
of the initial excitation energy remains captured near 
the bending region (see Fig. \ref{fig:scatt-1}c). 
In contrast, for $\kappa >2.6$ the main part of the 
excitation energy is reflected from the bending and only 
small part is captured (see Fig. \ref{fig:scatt-1}d).

\section{Conclusions}

In conclusion, we have shown that the interplay 
of curvature and nonlinearity in the chain of nonlinear
oscillators can lead to the qualitative effects, 
such as symmetry breaking of the nonlinear excitations 
and their trapping by the bending. 
We have demonstrated that the energy of nonlinear excitations
decreases with the increasing of curvature of the chain and 
thus one can expect that the presence of nonlinear localized
excitation in the chain may facilitate its bending, if only 
the chain is flexible enough. 

These effects are tolerant to the choice of the coupling 
coefficients $J_{n m}$. In particular, qualitatively the same 
properties were obtained for the case of the dipole-dipole 
dispersive interaction with 
$J_{n m} = |\vec{r}_n-\vec{r}_{m}|^{-3}$. 
Specifically, in the case of the 
Davydov-Scott model \cite{dav,scott} with the dipole-dipole 
coupling between sites \cite{leonor} for the parameters which 
characterize the motion of the amide-I excitation in proteins 
(they correspond to $N=0.64$ in our model) decreasing of the 
ground state energy due to a trapping by the bending with 
$\kappa=1$ ($\kappa=3$) comprise $10\%$ ($46\%$) of the whole 
excitation energy.
This may be also important for the dynamics of DNA molecule 
and in particular for DNA melting where in accordance with 
Peyrard-Bishop model \cite{pb,dpb} the nonlinear properties of
base-pairs vibrations play a crucial role.

\ack

We thank A.C.~Scott and M.~Peyrard for the helpful 
discussions.
Yu.B.G. and S.F.M. would like to express their thanks
to the Department of Mathematical Modelling, Technical
University of Denmark where the major part of this work
was done. S.F.M. acknowledges support from the 
European Commission RTN project LOCNET 
(HPRN-CT-1999-00163).



\end{document}